\documentclass[9pt,oneside,english]{amsart}
\usepackage[T1]{fontenc}
\usepackage[latin9]{inputenc}
\usepackage{geometry}
\usepackage{amstext}
\usepackage{amsthm}

\makeatletter
\numberwithin{equation}{section}
\numberwithin{figure}{section}
  \theoremstyle{remark}
  \newtheorem*{rem*}{\protect\remarkname}
  \theoremstyle{plain}
  \newtheorem*{conjecture*}{\protect\conjecturename}

\makeatother

\usepackage{babel}
  \providecommand{\conjecturename}{Conjecture}
  \providecommand{\remarkname}{Remark}

\begin{document}
\global\long\def\R{\mathbb{R}}

\global\long\def\S{\mathbb{S}}

\global\long\def\Z{\mathbb{Z}}

\global\long\def\C{\mathbb{C}}

\global\long\def\Q{\mathbb{Q}}

\global\long\def\N{\mathbb{N}}

\global\long\def\P{\mathbb{P}}

\global\long\def\F{\mathbb{F}}

\global\long\def\sgn{\text{sgn}}

\global\long\def\Stab{\text{Stab}_{\rho}}

\title{A Counterexample to the ``Majority is Least Stable'' Conjecture}

\author{Vishesh Jain}
\maketitle
\begin{rem*}
After this note appeared on the arXiv, we were informed that Sivakanth
Gopi (in 2013), and Steven Heilman and Daniel Kane (in 2017) already
independently observed that the ``Majority is Least Stable'' conjecture,
as stated, is not true. 
\end{rem*}
In their document Real Analysis in Computer Science: A Collection
of Open Problems, Filmus et al. \cite{key-2-1} present the following
``Majority is Least Stable'' conjecture due to Benjamini, Kalai
and Schramm \cite{key-1}: 
\begin{conjecture*}
Let $f\colon\{-1,1\}^{n}\rightarrow\{-1,1\}$ be a linear threshold
function for $n$ odd. Then, for all $\rho\in[0,1]$, $\Stab[f]\geq\Stab[\text{Maj}_{n}]$. 
\end{conjecture*}
In this note, we provide a simple counterexample to this conjecture,
even when $n=5$. We begin by noting that for an unbiased linear threshold
function $f:\{-1,1\}^{n}\rightarrow\{-1,1\}$, the statement $W^{1}[f]<W^{1}[\text{Maj}_{n}]$
would disprove the conjecture. An example of such an LTF is provided
by $f\colon\{-1,1\}^{5}\rightarrow\{-1,1\}$ given by 
\[
f(x_{1},x_{2},x_{3},x_{4},x_{5})=\sgn(2x_{1}+2x_{2}+x_{3}+x_{4}+x_{5})
\]

To show this, we explicitly compute the degree-1 Fourier coefficients
of $f$ and of $\text{Maj}_{5}$. Since both these functions are monotone,
computing $\hat{f}(i)$ is the same as computing the influence $\text{Inf}_{i}[f]$.
Moreover, by symmetry, it suffices to compute $\text{Inf}_{1}[\text{Maj}_{5}],\text{Inf}_{1}[f]$
and $\text{Inf}_{3}[f]$. 

Since coordinate $1$ is influential for $\text{Maj}_{5}$ iff $x_{2}+x_{3}+x_{4}+x_{5}=0$,
it follows that 
\[
\text{Inf}_{1}[\text{Maj}_{5}]=\frac{{4 \choose 2}}{2^{4}}=\frac{3}{8}
\]

Since coordinate $1$ is influential for $f$ iff $|2x_{2}+x_{3}+x_{4}+x_{5}|=1$,
and this happens precisely either if $x_{2}=\pm1$ and $x_{3}=x_{4}=x_{5}=\mp1$,
or $x_{2}=x_{i}=\pm1$, $x_{k},x_{l}=\mp1$ where $(i,j,k)$ is some
permutation of $(3,4,5)$, we get 
\[
\text{Inf}_{1}[f]=\frac{2+2\times{3 \choose 1}}{16}=\frac{1}{2}
\]

Finally, coordinate $3$ is influential for $f$ iff $2x_{1}+2x_{2}+x_{4}+x_{5}=0$.
This can happen iff $x_{1}=-x_{2}$ and $x_{4}=-x_{5}$. Hence
\[
\text{Inf}_{3}[f]=\frac{2\times2}{16}=\frac{1}{4}
\]

It follows that 
\begin{eqnarray*}
W^{1}[\text{Maj}_{5}] & = & \sum_{i=1}^{5}\text{\ensuremath{\hat{Maj_{5}}}}(i)^{2}\\
 & = & \frac{45}{64}
\end{eqnarray*}

whereas 
\begin{eqnarray*}
W^{1}[f] & = & \sum_{i=1}^{5}\hat{f}(i)^{2}\\
 & = & 2\times\left(\frac{1}{2}\right)^{2}+3\times\left(\frac{1}{4}\right)^{2}\\
 & = & \frac{44}{64}
\end{eqnarray*}

which completes the verification. 

\section*{Acknowledgements}

The author would like to thank Elchanan Mossel for help with the presentation
of the argument, as well as introducing him to \cite{key-2-1}.

\end{document}